\documentclass[conference]{IEEEtran}
\IEEEoverridecommandlockouts

\usepackage{amsmath,amssymb}
\usepackage{algorithmic}
\usepackage{booktabs}
\usepackage{threeparttable}
\usepackage{multirow}
\usepackage{svg}
\usepackage{graphicx}
\usepackage{rotfloat}
\usepackage{float}
\usepackage{stfloats}
\usepackage{textcomp}
\usepackage{xcolor}
\usepackage{caption}
\usepackage{biblatex}
\usepackage{footnotebackref}[2012/07/01]
\usepackage{tablefootnote}[2014/01/26]
\bibliography{ref_bib}

\begin{document}

\title{A Survey of Side-Channel Attacks in Context of Cache - Taxonomies, Analysis and Mitigation}

\author{
\IEEEauthorblockN{ 
Ankit Pulkit, Smita Naval, Vijay Laxmi}
\IEEEauthorblockA{\textit{Dept. of Computer Science and Engineering}\\
\textit{Malaviya National Institute of Technology Jaipur}\\
India \\
2020rcp9585@mnit.ac.in, smita.cse@mnit.ac.in,vlaxmi@mnit.ac.in}
}
\maketitle

\begin{abstract}
Nowadays, side-channel attacks have become prominent attack surfaces in cyberspace. Attackers use the side information generated by the system while performing a task. Among the various side-channel attacks, cache side-channel attacks are leading as there has been an enormous growth in cache memory size in last decade, especially Last Level Cache~(LLC). The adversary infers the information from the observable behavior of shared cache memory. This paper covers the detailed study of cache side-channel attacks and compares different microarchitectures in the context of side-channel attacks. Our main contributions are: (1) We have summarized the fundamentals and essentials of side-channel attacks and various attack surfaces. We also discussed different exploitation techniques, highlighting their capabilities and limitations. (2) We discussed cache side-channel attacks and analyzed the existing literature on cache side-channel attacks on various parameters like microarchitectures, cross-core exploitation, methodology, target, etc. (3) We discussed the detailed analysis of the existing mitigation strategies to prevent cache side-channel attacks. The analysis includes hardware- and software-based countermeasures, examining their strengths and weaknesses. We also discussed the challenges and trade-offs associated with mitigation strategies. This survey is supposed to provide a deeper understanding of the threats posed by these attacks to the research community with valuable insights into effective defense mechanisms.
\end{abstract}

\section{Introduction}
The world jumped into the Digital Revolution or the Third Industrial Revolution in the late 20$^{th}$ century \cite{digital_era1,digital_era2}. In the pre-digital era, paper files (or mechanical and analog electronic technologies) dominated, with digital-era computers predominated to store and process information. Because of this revolution, overall information is present in cyberspace for now. This revolution triggered intended or unintended disclosure of information and unauthorized access events \cite{info_cyber1,info_cyber3,info_cyber2}. Though computer security policies were there to protect the confidentiality of the data, most of the policies emphasized role-based access control, access control matrix, and the chain of trust \cite{comp_sec_policy}. Initially, it was enough to provide security to the confidential data, but it could not defend against side-channel attacks because adversaries did not gather the information by exploiting direct access.

The increasing adoption of shared cache architectures in modern processors, driven by the demand for performance optimization, has inadvertently introduced side and covert channels that attackers can exploit. Shared caches store frequently accessed data closer to the processor cores, accelerating data retrieval and reducing memory latency \cite{adaption_of_cache}. However, while enhancing system performance, this architectural optimization also creates a potential vulnerability that malicious entities can exploit \cite{cache_vul_2,cache_vul_3,cache_vul_4,cache_vul_1}. Nowadays, the computer architecture community has significantly tried to adapt the methodologies to protect against side-channel attacks. Researchers also worked extensively on side-channel attacks and their mitigation methodologies in the last decade.

In side-channel attacks, the attacker process uses the some shared resource to track the victim process. The attacker receives data from that shared resource, such as timing information (like the time an operation takes to complete), etc. The attacker uses this observed data, as a result of which he converts it into information. As the demand for computer systems is increasing, scientists are paying more attention to using shared resources to utilize the available resources fully. Furthermore, this caused a higher risk of side-channel attacks.

Paul C. Kocher \cite{ref_kocher_96,ref_kocher_dpa} introduced timing channels in 1996; he demonstrated how timing measurements of a system performing a cryptographic function could be used to find the secret key of various cryptosystems like Diffie-Hellman, DSS, RSA, and other systems. His work includes the techniques used in timing attacks, statistical analysis, and deriving the correlation between execution time and secret key to deduce the secret key. Later, researchers started thinking in various directions with the concept that a measured quantity can lead to information leakage, and this is how various side-channels \cite{researcher6,researcher8,researcher13,researcher4,researcher16,researcher15,researcher12,researcher5,researcher14,researcher7,researcher3,researcher11,researcher1,researcher2,researcher9,tab30} were introduced. Initially, researchers focused only on compromising cryptographic algorithms and devices. However, as the topic reached the boom, people started working on side-channel attacks with a scope that crossed the cryptographic-level border. Now many researchers are working on exploiting areas other than the cryptography level.
These \cite{noncrypt2,blackhat2016,noncrypt3} side-channel attacks are non-cryptographic attacks.

To comprehensively address cache side channel attacks, it is essential to understand the broader context of vulnerability, cache memory and side channel attacks. We discussed vulnerabilities with the reference of the CVE database of the MITRE corporation.
Side channel attacks provide a range of techniques that can exploit unintended information leakages, such as timing, power consumption, electromagnetic emanations, and acoustic emissions. We can understand the unique characteristics and repercussions of cache side channel attacks by analyzing these attack vectors.

Our main aim in writing this paper is to provide an in-depth overview of cache side-channel attacks, covering their fundamentals, attack models, and detection techniques. We address the origin and evolution of cache side-channel attacks and explore the complex mechanisms underlying these attacks. By examining basic concepts of cache memory organization, such as cache hierarchy, replacement policy, and coherence protocol, we lay the foundation for understanding the vulnerabilities exploited by attackers.

We discuss prominent cache side-channel attack scenarios to enhance understanding, including Prime + Probe \cite{tab37,prime_probe1}, Flush + Reload \cite{flush_reload2,flush_reload1,researcher9}, Evict + Time \cite{tab8,evict_time1}, Evict + Reload,  and Flush + Flush \cite{flush_flush1,tab39}. These scenarios illustrate how attackers can leverage cache behavior to infer sensitive information. By examining their working principles and discussing real-world implications, we comprehensively understand the potential threats of cache side-channel attacks.
Furthermore, we conduct a detailed analysis of existing mitigation techniques to prevent cache side-channel attacks. These countermeasure techniques include hardware and software-based countermeasures with their strengths and limitations. Hardware-based techniques include cache partitioning, where cache resources are isolated between security domains, and cache obfuscation, which introduces noise and confusion to trick the attackers. On the other hand, software-based approaches have compiler optimizations, code randomization, and cryptographic techniques.

While discussing the effectiveness of countermeasures, we also address the challenges and trade-offs associated with cache side-channel attack mitigation. Countermeasures often introduce performance overhead, impact energy consumption, and may complicate the software development process. Achieving a balance between security and system efficiency is crucial to ensure practical and sustainable implementations of mitigation strategies. In addition to the current state of cache side-channel attacks and mitigation techniques, this survey paper explores the evolving landscape of these attacks. We discuss emerging attack vectors and potential future research directions, enabling researchers and practitioners to develop robust defense mechanisms proactively. By identifying vulnerabilities and addressing emerging threats, we aim to contribute to developing resilient computer systems that can effectively thwart cache side-channel attacks while maintaining optimal performance and usability.

In summary, this survey paper provides a comprehensive and insightful examination of cache side-channel attacks, including their source, mechanisms, and detection techniques. By analyzing existing mitigation strategies and discussing their challenges, we aim to encourage a deeper understanding of the threats posed by cache side-channel attacks and facilitate the development of effective defense mechanisms. Ultimately, the goal is to enhance the security and stability of computer systems in the face of cache side-channel attacks, striking a balance between protection and performance.

\subsection{Related Surveys}

\begin{table}[H]
    \centering
    \begin{tabular}{cp{1cm}cp{1cm}p{3.3cm}}
        \toprule[0.6mm]
        \textbf{Year} & \textbf{Survey} & \textbf{Target} & \textbf{Topics} & \textbf{Highlights} \\  \addlinespace[0.2em] \toprule 
         2016 & Szefer \cite{survey_micro2} & Cloud & MSCA\makebox[0pt][l]{$^{\ast}$} & Cache side-channel attacks are part of discussed microarchitectural attacks.Considered various attacks on L1, L2, and LLC Caches.\\
         \midrule 
         2017 & Biswas et al. \cite{survey_timing} & H/W & Timing SCA & Timing cache side-channel attacks only.\\
         \midrule 
        2017 & Lyu et al. \cite{survey_cache2} & IoT  & Cache SCA & Cache side-channel attacks mainly on Internet of Things (IoT). Classification and countermeasures.\\
        \midrule 
        2018 & Ge et al. \cite{survey_micro1} & H/W & MSCA\makebox[0pt][l]{$^{\ast}$} &  Cache side-channels on shared cache architectures, classification, and countermeasures \\  \midrule 
        2018 & Spreitzer et al. \cite{survey_mobile_device} & H/W & Mobile Devices & Various side-channel attacks on Mobile and their classification \\ \midrule 
        2019 & Mirzargar et al. \cite{survey_physical} & H/W & FPGA SCA &  Various physical side-channel attacks son FPGAs and classification. \\ \midrule 
        2021 & Lou et al. \cite{noncrypt3} & Crypto & MSCA\makebox[0pt][l]{$^{\ast}$} & Microarchitectural side-channel attack on cryptosystems and countermeasures\\ \midrule 
        2023 & Our & System & Cache SCA & Essential concepts, methodologies of side-channel attacks,and  the taxonomy of these attacks. Cache side-channel attacks, their types, analysis, and countermeasures. \\ 
        \bottomrule[0.6mm]

    \end{tabular}
    \begin{tablenotes}
        \item \footnotesize$^*$MSCA: Microarchitectural Side-Channel Attacks
    \end{tablenotes}
    \caption{Related surveys and their highlights in brief}
    \label{tab:my_table_related_survey}
\end{table}

In this survey paper, we have conducted a comprehensive survey on cache side-channel attacks. There have been several surveys on cache side-channel attacks in past years, each with its focus and strengths. We have discussed a brief comparision in TABLE \ref{tab:my_table_related_survey}. The following details are on the mentioned surveys.

The survey by Szefer \cite{survey_micro2} discusses the microarchitectural side and covert channel attacks. Cache side-channel attacks are part of the discussed microarchitectural attacks. The author categorizes attacks based on virtualization and highlights their channel bandwidth. Considered various attacks on L1, L2, and LLC Caches.

The survey by Biswas et al. \cite{survey_timing} comprehensively discussed timing channel attacks and focused on a few cache side-channel attacks. Considered only timing cache side-channel attacks, it does not consider trace and access-driven-based cache side-channel attacks.

Another survey by Lyu et al. \cite{survey_cache2} considered various cache side-channel attacks, mainly with the Internet of Things (IoT). The authors discussed various categorizations of attacks like attacks on single-core systems, multi-core systems, and cross-vm attacks. Various countermeasures were also discussed in this survey.

The survey by Ge et al.  \cite{survey_micro1} considered various microarchitectural attacks, including several cache side-channel attacks. The authors emphasized side-channels on shared cache architectures and classified attacks based on shared resources. The authors focused on virtualized environment attacks especially.

Another survey by Spreitzer et al. \cite{survey_mobile_device} considered various side-channel attacks, and the paper's main contribution is to classify the side-channel attacks based on active and passive classes. The authors only considered side-channel attacks on mobile devices. We extended this to overall side-channel attacks.

The survey by Mirzargar et al. \cite{survey_physical} surveyed physical side-channel attacks, especially on FPGAs, and classified these side- and covert-channel attacks. We considered software-based side-channel attacks mainly.

The survey by Lou et al. \cite{noncrypt3} discussed various microarchitectural side-channel attacks. The authors emphasized various cryptosystems and their weaknesses against these attacks. Cache side-channel attacks were included as a primary portion of this survey. Various countermeasures were also discussed.

Our survey discusses the essential concepts and methodologies for side-channel attacks and the taxonomy of these attacks. We emphasized cache side-channel attacks, mainly including cache architecture, different methodologies, analysis of various attacks, and present existing countermeasures.

\section{Terminologies}
This section discusses some concepts required for side-channel attacks briefly. Topics include an overview of vulnerabilities, processor architecture, and other technologies. The TABLE \ref{tab:my_table_abbreviations} has a list of abbreviations used in various sections of the paper.
\begin{table}[h]
    \centering
    \begin{tabular}{p{2cm}p{6cm}}
        \toprule[0.6mm]  
        \textbf{Abbreviations}& \textbf{Definition} \\ \addlinespace[0.2em] \toprule 
        AES 	&	Advanced Encryption Standard	\\	\midrule
        ASLR 	&	Address Space Layout Randomization	\\	\midrule
        CPU 	&	Central Processing Unit	\\	\midrule
        CRT	    &	Cathode-Ray Tube	\\	\midrule
        CT      &   Cipher Text \\ \midrule
        CVE 	&	Common Vulnerabilities and Exposures	\\	\midrule
        CVVS    &   Common Vulnerability Scoring System \\ \midrule
        DSS 	&	Digital Signature Scheme	\\	\midrule
        DTLS	&	Datagram Transport Leyer Security	\\	\midrule
        EM  	&	Electro Magnetic	\\	\midrule
        KASLR 	&	Kernel Address Space Layout Randomization	\\	\midrule
        L1	&	Level 1 Cache	\\	\midrule
        L1-D 	&	Level 1 - Data Cache	\\	\midrule
        L1-I 	&	Level 1 - Instruction Cache	\\	\midrule
        L2	&	Level 2 Cache	\\	\midrule
        L3	&	Level 3 Cache	\\	\midrule
        LLC 	&	Last Level Cache	\\	\midrule
        MB	&	MegaByte	\\	\midrule
        NIST 	&	National Institute of Standards and Technology	\\	\midrule
        NSA 	&	National Security Agency	\\	\midrule
        NVD     &   National Vulnerability Database \\ \midrule
        PT 	&	Plain Text	\\	\midrule
        PQC 	&	Post Quantum Cryptography	\\	\midrule
        RAM	&	Random Access Memory	\\	\midrule
        RSA 	&	Rivest–Shamir–Adleman	\\	\midrule
        SMT 	&	Simultaneous Multithreading	\\	\midrule
        SSL	&	Secure Socket Layer	\\	\midrule
        TLS	&	Transport Layer Security	\\	\midrule
        VM	&	Virtual Machine	\\	\midrule
        WWII 	&	World War II	\\	\bottomrule[0.6mm]
    \end{tabular}
    \caption{List of abbreviations used in this paper}
    \label{tab:my_table_abbreviations}
\end{table}
\subsection{Vulnerability}

Vulnerability in computer security refers to a weakness or flaw in a computer system or software that can be exploited by malicious actors to gain unauthorized access, steal data, or cause other types of harm. These vulnerabilities can arise due to programming errors, configuration issues, or even the use of outdated or unsupported software. 

Hackers and other cyber criminals often use automated tools to scan for known vulnerabilities and launch attacks, making it essential for organizations to regularly patch and update their systems to address these issues. Failure to address vulnerabilities can result in data breaches, financial losses, and damage to an organization's reputation. Therefore, identifying and addressing vulnerabilities is a critical component of effective computer security.

Vulnerability refers to a weakness or flaw in a system or software that can be exploited by a threat actor to carry out an attack \cite{ref_fips_vul,vulnerability_def}. A threat, on the other hand, is an event or action that has the potential to cause harm to a system or organization. Threats can come from a variety of sources, including cybercriminals, insiders, natural disasters, and human error. Risk, in the context of cybersecurity, is the likelihood of a threat exploiting a vulnerability and causing harm to an organization. It is the product of the probability of a threat occurring and the impact that it would have on the organization. Therefore, risk can be seen as a measure of the potential damage that could be caused by a vulnerability being exploited by a threat.
\begin{figure}[H]
    \centering
    \includegraphics[scale=0.65]{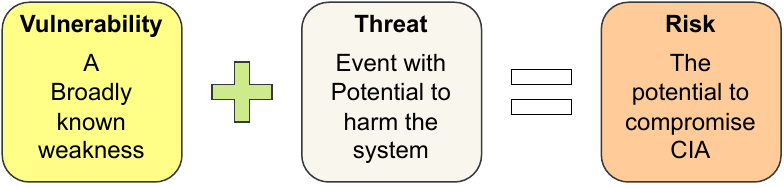}
    \caption{Vulnerability and Associated Risk(s)}
    \label{fig:my_label_vul_threat_risk}
\end{figure}

\begin{table*}[ht]
    \centering
    \resizebox{\textwidth}{!}{
    \begin{tabular}{cp{7.8cm}cp{1.5cm}p{2.7cm}}
     \toprule[0.6mm]
        \textbf{CVE ID\textsuperscript{$\ast$}} & \textbf{Description} & \textbf{Score\textsuperscript{$\dagger$}} &\textbf{Exploit Availability} &\textbf{Domain}  \\ \addlinespace[0.2em] \toprule 
         CVE-2023-5326 & A critical vulnerability found in SATO CL4NX-J Plus 1.13.2-u455\_r2 which affected by an unknown functionality of WebConfig component. This leads to improper authentication on local network. & 8.8 & Public & Network Vulnerability \\ \midrule
         CVE-2023-3090 & The Linux Kernel's ipvlan network driver has a heap out-of-bounds write vulnerability. This can be used to get control over the system with local privilege escalation & 7.8 & Public &Operating System Vulnerability \\ \midrule
         CVE-2023-5053 & A bypass authentication vulnerability found in Hospital Management System (v378c157) as application is vulnerable to SQL Injection. & 9.8 & Public & Application Vulnerability \\ \midrule
         CVE-2022-42703 & \textit{mm/rmap.c} in the Linux kernel before 5.19.7 has a use-after-free related to leaf anon\_vma double reuse. &5.5 & Public & Kernel Vulnerability \\ \midrule
         CVE-2022-36619 & In D-link DIR-816 A2\_v1.10CNB04.img,the network can be reset without authentication via \textit{/goform/setMAC}.  & 7.5 & Public & Network Vulnerability \\ \midrule
         CVE-2021-30688 & A malicious application may to break out of its sandbox. This issue is fixed in \textit{macOS Big Sur 11.4}, Security Update 2021-003 Catalina. A path handling issue was addressed with improved validation. & 8.8 & Not Available & Application Vulnerability\\ \bottomrule[0.6mm]
    \end{tabular}}
    \begin{tablenotes}
        \item \textsuperscript{$\ast$} This ID is given to identify, define, and catalog publicly disclosed cybersecurity vulnerabilities by MITRE Corporation.
        \item \textsuperscript{$\dagger$} It is NVD Vulnerability Severity Base Score computed by NIST.
    \end{tablenotes}
    \caption{List of Vulnerabilities}
    \label{tab:list_vul}
\end{table*}

\begin{table}[]
    \centering
    \resizebox{\linewidth}{!}{
    \begin{tabular}{lclc}
        \toprule[0.6mm]
        \multicolumn{2}{c}{ \textbf{CVSS v2.0 Ratings}}  & \multicolumn{2}{c}{\textbf{CVSS v3.0 Ratings}} \\ \cmidrule(lr){1-2} \cmidrule(lr){3-4}
        
        \multicolumn{1}{c}{  {  \textbf{Severity}}} & \multicolumn{1}{c}{  \textbf{Base   Score Range}} & \multicolumn{1}{c}{  \textbf{Severity}} & \multicolumn{1}{c}{ \textbf{Base   Score Range}} \\  \addlinespace[0.2em] \toprule
        & {  }& {None}& {0} \\ \midrule
        {Low} & {  0.0-3.9}& {Low} & {  0.1-3.9} \\ \midrule
        {Medium}& {  4.0-6.9}& {Medium} & {  4.0-6.9}\\ \midrule
        {High}& {  7.0-10.0}& {High}& {  7.0-8.9} \\ \midrule
        { }& {  }& {Critical}& {  9.0-10.0}  \\ \bottomrule[0.6mm]                                              
    \end{tabular}}
    \caption{NVD Vulnerability Severity Ratings}
    \label{tab:vul_score_risk}
\end{table}
Vulnerability, threat, and risk are interconnected concepts that are fundamental to understanding the cybersecurity landscape as shown in Fig. \ref{fig:my_label_vul_threat_risk}. The relationship between vulnerability, threat, and risk is such that vulnerabilities create opportunities for threats, which in turn increase the risk to an organization \cite{cyber_threat_vuln1,cyber_threat_vuln}. As such, cybersecurity professionals must identify vulnerabilities in their systems and software, assess the likelihood and potential impact of threats, and take measures to mitigate the risks associated with those threats. This may involve implementing security controls, updating software, training employees, and regularly assessing the security posture of the organization. By understanding and managing the relationship between vulnerability, threat, and risk, organizations can better protect themselves against cybersecurity threats and minimize the potential impact of an attack.

In other words, vulnerability is a weakness of a computer system that can allow an attacker to gain unauthorized privileges if exploited. When a vulnerability and a threat are encountered together (A threat is an event that can potentially exploit a vulnerability), it may cause direct or indirect risk to the computer system. TABLE \ref{tab:list_vul} shows some recent vulnerabilities of various types. The column \textit{Score} in TABLE \ref{tab:list_vul} indicates the severity score of a particular vulnerability computed by NIST. To map the severity with the score TABLE \ref{tab:vul_score_risk} is helpful. The column \textit{Exploit Availability} in TABLE \ref{tab:list_vul} indicates if any exploit related to a particular vulnerability is available in public domain or not. The column \textit{Domain} may vary based on the capability of the vulnerability. The affected area of the computer system on its exploitation decides the capability of the vulnerability. Side-Channel vulnerabilities are explained in the another section. 
\subsection{Signal to noise ratio (SNR)}
\subsubsection{Signal}
In the field of communication, a signal is the actual information that we want to gain. It is the part of data that we receive from another endpoint. The receiver needs to process the data to interpret the required information (signal). A signal may be an electric waveform, audio, video, or some other form of data containing the required information. In digital communications, a signal is a sequence of bits.
\subsubsection{Noise}
Noise is the unwanted information or random disturbance attached to the signal while receiving the data. There can be multiple reasons for noise in the data, like atmospheric conditions, equipment misconfiguration, electric interference, magnetic interference, etc. Noise may be added at various phases, like recording, processing, and transmitting the data. Heavy noise can manipulate the original signal significantly.
\subsubsection{Signal to Noise Ration (SNR)}
It is a measure of the ratio of the strength of the desired signal and the level of the noise. It is considered essential to understand the quality of the transmission. If $P_{signal}$ is the power of signal (desired information) and $P_{noise} $ is the power of noise (unwanted information) then:\\

\begin{center}
    $ SNR = \dfrac{P_{signal}}{P_{noise}} $    
\end{center}

\subsection{Simultaneous Multithreading (SMT)}
\begin{figure}
    \centering
    \includegraphics[scale=0.9]{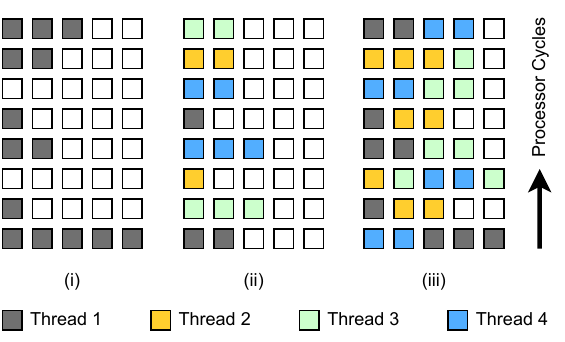}
    \caption{Simultaneous Multithreading}
    \label{fig:smt}
\end{figure}

Simultaneous Multithreading (SMT) is an architectural technique developed to improve the processor`s efficiency using the concept of hardware multithreading in modern microprocessor design \cite{smt_ref}. It allows a physical core to run multiple threads simultaneously. SMT, also known as Hyper-Threading in Intel`s processors, helps a physical core share its resources among multiple threads like pipelines to process instructions, cache memory, etc. It boosts the processor`s performance by improving the overall utilization and throughput as it is visible in Fig. \ref{fig:smt}. However, there is a security risk along with the boosted performance, such as side-channel vulnerabilities. 

Considering SMT, the simultaneous execution of multiple threads may lead to sharing the data or state among the inter-process threads. These intended or unintended channels might lead to cache side-channel attacks and branch prediction attacks to infer sensitive information \cite{noncrypt3}. Information security research communities and industrial experts presented countermeasures, including fine-grained control over shared resources, cache partitioning methods, etc. As the landscape of side-channel attacks increases, attackers try exploiting novel vulnerabilities using simultaneous multithreading, so security remains an area of interest. 

\subsection{Cryptography}
Cryptography is useful to secure sensitive information and communication to ensure confidentiality, integrity, and authentication. The main purpose of cryptography is to protect our sensitive information from unauthorized access and manipulation of the information over transmission. Considering side-channel attacks as an indirect approach to invading a system's security, cryptography might be a victim in the case of side-channel attacks. In cryptography, various algorithms use various operations (logical and mathematical) in encryption and decryption. By observing the unintended or intended information, an attacker might get access to the key used in the cryptographic algorithms. That leads to various severe security concerns. These side-channel attacks boost the throughput of brute-force and mathematical analysis approaches. Researchers worked on some countermeasures (masking, randomization, hardware techniques, etc.) against cryptographic side-channel attacks, but as side-channel attacks evolved, attackers came up with new devastating methods to carry out such attacks.

\section{CPU Cache Interface}
Cache memory, or cache, is a chip-level computer component with storage functionality. The physical location of cache memory is between the main memory and processor units. The cache memory's speed in performing reads and writes is very-very fast. However, when we consider the cache memory's capacity is restrictive to only a few MBs. Cache memory was introduced to improve the performance of the system. It has the same working as the main memory.

\begin{figure}[h]
    \centering
    \includegraphics[width=\linewidth]{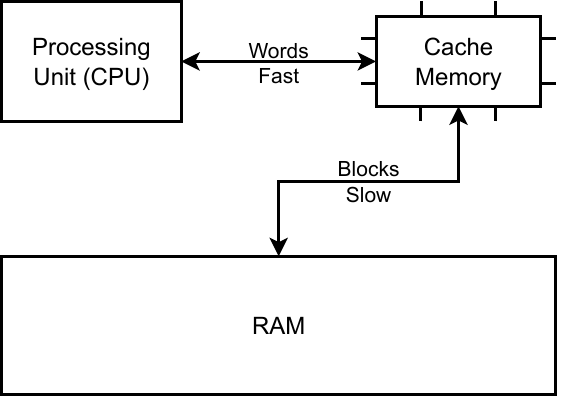}
    \caption{Single Level Cache}
    \label{fig:my_label_l1_cache}
\end{figure}

As shown in Fig. \ref{fig:my_label_l1_cache} (for single level cache-memory), when the processor requests the data/instruction, first it checks whether the data is available in cache memory or not, and if the data is unavailable in cache memory, then it moves towards main memory, and so on.\\ 
\begin{figure}[h]
    \centering
    \includegraphics[width=\linewidth]{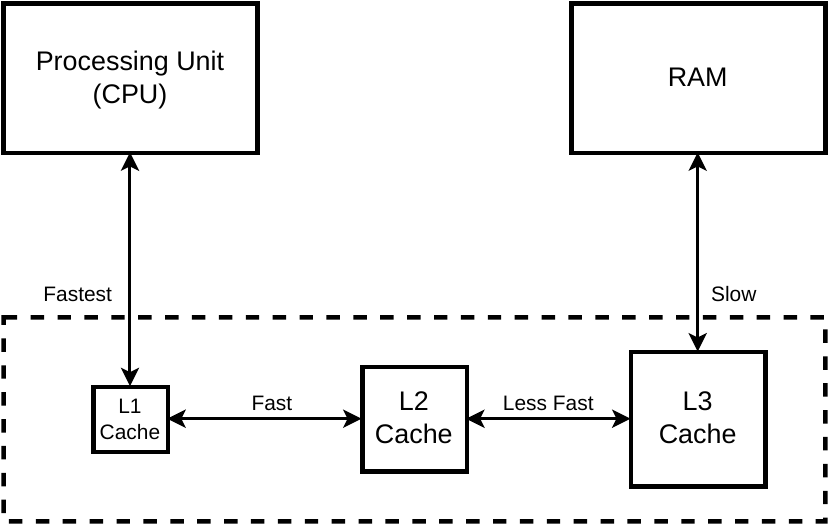}
    \caption{Multi Level Cache}
    \label{fig:my_label_l3_cache}
\end{figure}


Fig. \ref{fig:my_label_l3_cache} illustrates the processing for multi-level cache-memory; the request is entertained by the L1-cache (Level-1) when processor request for the data/instruction if the requested data/instruction is available in L1-cache then it is fine; otherwise, it searches from L2-cache and so on. Here the categorization of cache memory has been done based on storage capacity and access speed. L1-cache has less storage capacity but the fastest access, and L3-cache has higher storage capacity but the slowest access out of these three cache levels. L2-cache has higher storage and less access speed than L1-cache and less storage and higher access speed than L3-cache. If the data/instruction is available in L1-cache, it must be in L2-cache and L3-cache. TABLE \ref{tab:cache_list_table} shows various processor models and architecture has different cache memories. The size of L1, L2, and L3 cache memories differs for available processors in the market. It depends on the manufacturers and the architecture. 
\begin{table}[h]
  \centering
  \resizebox{\linewidth}{!}{
    \begin{tabular}{lcccc}
    \toprule[0.6mm]
         \textbf{Model} & \textbf{Microarchitecture} & \textbf{L1\textsuperscript{$\ast$}} & \textbf{L2\textsuperscript{$\ast$}} & \textbf{L3\textsuperscript{$\dagger$}}  \\ \addlinespace[0.2em] \toprule
         i3-4005U & Haswell ULT & 128 & 512 & 3 \\ \midrule
         AMD Ryzen 3 3250U & Picasso & 192 & 1024 & 4 \\ \midrule
         i7-8750H  & Coffee Lake  & 384 & 1536 & 9 \\ \midrule
         AMD Ryzen 7 6800H  & Rembrandt  & 512 & 4096 & 16 \\ \midrule
         i7-12700  & Alder Lake & 960 & 11,468 & 25 \\ \bottomrule[0.6mm]
    \end{tabular}}
    \begin{tablenotes}
      \item \textsuperscript{$\ast$}Capacity in KBs
      \item \textsuperscript{$\dagger$}Capacity in MBs
    \end{tablenotes}
  \caption{Cache memory size in various systems}
  \label{tab:cache_list_table}
\end{table}
When the data (or instructions) exists in the cache memory, it is said cache hit; otherwise, it is named cache miss. In case of a cache miss, the data move from the main memory to the cache memory, which the processor then accesses. A concept of reusing the particular data or data used recently, because a cache is directly attached to the processor unit, creates a high probability of reusing the cache. Thus it minimizes the CPU's access time and improves the system's overall efficiency.

\section{Side-Channel Attacks}
\label{sec:side_channel_attack_history}

Let us take an example of a vault where diamonds or many valuables are stored. Please assume the vault has a 6-digit pattern to open its mechanical lock. Then, there are 1,00,000 possible patterns to open the vault. If an attacker tries to break the vault’s mechanical lock with the brute force technique, at least 2/3 of all the possible patterns need to be applied to get the correct number ideally. It can take several days, so the vault is considered secure. However, when a lock picker tries the patterns and observes the rotor motion or the sound produced when the rotors are set to the particular digit, the lock picker opens the vault in a few minutes. This is a perfect example of side-channel attacks.

According to Corina S. et al., 2019 ``Side-Channels allow an attacker to infer information about a secret by observing nonfunctional characteristics of a program, such as execution time or memory consumed.\cite{def_1}'' Or Paul Grassi et al., states, ``An attack enabled by information leakage from a physical cryptosystem. Characteristics that could be exploited in a side-channel attack include timing, power consumption, and electromagnetic and acoustic emissions\cite{def_2}." The first definition covers widespread attacks, including cryptographic and non-cryptographic attacks, whereas the second only empowers cryptanalysis-based side-channel attacks.
\begin{figure}[h]
    \centering
    \includegraphics[scale=1.35]{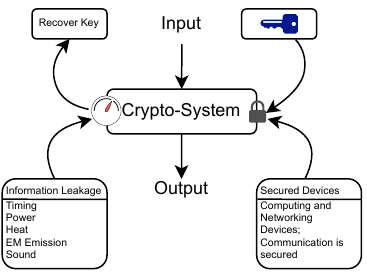}
    \caption{Side-Channel Attack}
    \label{fig:my_label_side_channel_basic}
\end{figure}
Fig. \ref{fig:my_label_side_channel_basic} explains that the side-channel has nothing to do with direct access to the confidential data, which is secure. In a side-channel attack, an attacker observes the leakage of nonfunctional characteristics such as timing, power, heat, EM emission, etc., later by training the model or comparing it with some known data is changed into the data the victim wanted to protect.

Many articles \cite{ref_sca_web2,ref_sca_web1,ref_sca_web3,ref_sca_web4} claim that side-channel attacks used during the World War II and they show it as the earliest example. An approved document\cite{ref_tempest1} for release by NSA in 2007 in response to the Freedom of Information Act (FOIA) request to the Department of Defense, USA. The document's title was ``TEMPEST: A Signal Problem" Initially, the document was listed under the secret/classified category but was later released in the public domain by the NSA of the USA.
\begin{figure}[h]
    \centering
    \includegraphics[width=\linewidth]{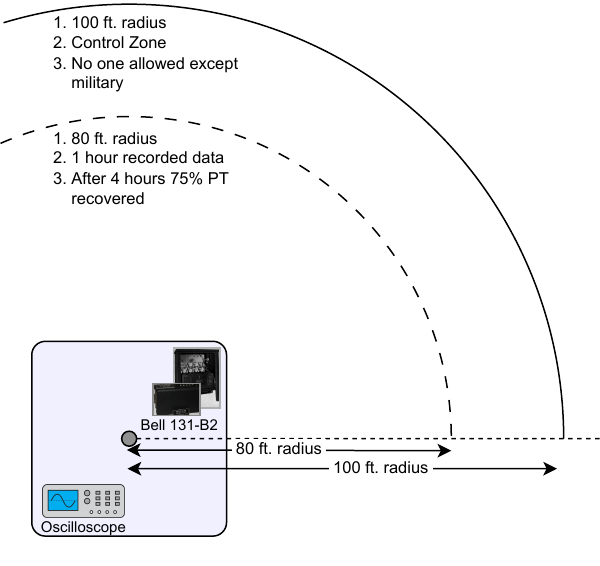}
    \caption{TEMPEST Scenario}
    \label{fig:my_label_WWII}
\end{figure}
Bell 131-B2 mixing devices were used for secured/encrypted communication during WWII. The researcher(s) of Bell Labs noticed that an electromagnetic spike was seen with the help of an oscilloscope when the mixing machine stepped. It happened because of rotors rotating/shifting. Later, the researchers concluded that plain text could be recovered from the cipher text. Later this experiment was carried out at a large scale, and Signal Corps found that approximately 3/4 of the original text could be recovered using the mentioned technique, which was 80 feet away. The above mentioned scenario is explained in Fig. \ref{fig:my_label_WWII} \\
The document also states that three methods were used to suppress the attack:
\begin{itemize}
    \item Shielding: 
    They covered the machine with multiple layers of various materials to prevent EM radiation.
    \item Filtering: 
    Filters were installed to stop the electromagnetic radiations.
    \item Masking:
    They used multiple layers of fake signals to add noise to the original signal of the machine by assuming that it would protect their original signal. 
\end{itemize}

Though they used all three mentioned techniques, a modified version could not be placed because shifting all the machines to the Bell Labs was impossible. Moreover, using the mentioned techniques took much work to implement, as it introduced various operational issues like heat generated by the machine, etc.. Hence, the Signal Corps ordered the commanders to secure a radius of 100 feet.

If we consider the overall side-channel category, then a research paper was published under the unclassified category titled “Electromagnetic Radiation from Video Display Units: An Eavesdropping Risk?”  by Wim Van Eck\cite{ref_EMradiation}. Eck demonstrated eavesdropping on the CRT (Cathode Ray Tube) video display unit in this paper. Initially, he realized that radiation measurement was not synchronized, so images and pixels moved horizontally and vertically simultaneously. Later, with the help of a programmable digital frequency divider, WV Eck could reconstruct both horizontal and vertical synchronizations. He also proposed some solutions to avoid such attacks. Provided solutions include minimization of radiation level, maximizing the noise, and cryptographic display.

Paul C. Kocher published timing measurement or attack for the first time in Year 1996\cite{ref_kocher_dpa}. The Base idea of Kocher was to measure the time each process takes in the critical section. According to Kocher, each operation requires a different amount of time to execute, or it can be stated as each operation takes different times to get executed. This was the key difference; Kocher took a bet. He measured the execution of a process with various inputs and successfully broke Diffie-Hellman, RSA, and some other cryptographic systems.

So the basic concept of side-channel attacks is to get the information by observing the process, hardware or any other component through the same device or sometimes external devices. Furthermore, exploit the observed data to deduce the information as seen in Fig. \ref{fig:my_label_side_channel_basic}.
\subsection{Road-map to the Side-Channel Attack(s)}
A side-channel attack is completely different from traditional attacks. Some attacks might require the physical access of the system. Due to this being different from other attacks, we must follow the steps below to carry out this attack successfully. 

\subsubsection{Reconnaissance}
Initially, the attacker gathers information about the target system or device, including its hardware and software configurations, use cases, and potential vulnerabilities.
\subsubsection{Profiling} 
In this stage, the attacker collects baseline data about the target system or device, including its power consumption, timing behavior, electromagnetic emissions, or other side-channel characteristics.
\subsubsection{Attack planning}
In this stage, the attacker selects the appropriate side-channel attack technique based on the characteristics of the target system or device, as well as the level of access and resources available.
\subsubsection{Attack execution}
In this stage, the attacker executes the selected side-channel attack technique to infer information about the victim's data or secret key. The attacker may use a combination of hardware and software tools to perform the attack, and may need to adapt their approach based on the results of the profiling stage.
\subsubsection{Data analysis}
In this stage, the attacker analyzes the data collected during the attack to extract sensitive information about the victim's data or secret key. This may involve statistical analysis, machine learning algorithms, or other advanced techniques to extract meaningful patterns or correlations from the data.
\subsubsection{Exploitation}
In this stage, the attacker uses the information obtained from the side-channel attack to achieve their objectives, which may include accessing restricted data, impersonating the victim, or compromising the security of the system or device.
\subsubsection{Covering tracks} In this stage, the attacker covers their tracks by removing any traces of the side-channel attack, including the data collected, the tools used, and any modifications made to the target system or device.

\section{Taxonomy and Working of Side-Channel Attacks}

\begin{figure}
    \centering
    \includegraphics[scale=0.7]{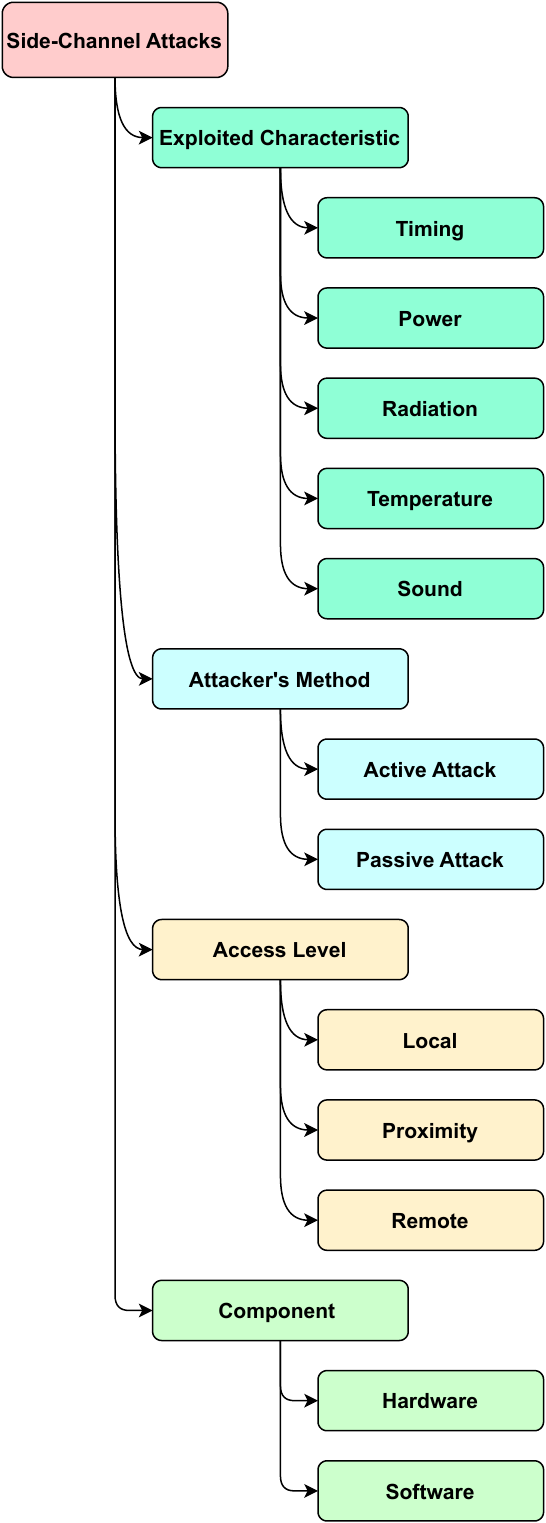}           
    \caption{A general taxonomy of side-channel attacks }
    \label{fig:general-taxonomy}
\end{figure}

Side-channel attacks can be classified into different types based on various parameters, such as the type of information being leaked, the channels used for leakage, the software or hardware component involved, and whether the execution environment is modified. Fig. \ref{fig:general-taxonomy} presents a general classification of side-channel attacks, classified according to various criteria.\\
The considered parameters to provide a coarse-grained classification of side-channel attacks are \textit{Exploited characteristics}, \textit{Attacker`s method}, \textit{Access level}, and \textit{Component}. 

The term \textit{exploited characteristics} indicates the unique characteristic of a system, whether physical or logical, that shows a relative behavior with an ongoing operation or with the data involved in some process. The characteristic could be anything that can be measured to gain information about the system, such as timing, power consumption, or other measurable characteristic. 
The second parameter \textit{attacker's method} focuses on the method attacker's invasion in the victim's system. Whether it involved some interference with the execution environment of the victim's system or the attacker merely observes some characteristics.The third parameter \textit{access level} considers three aspects based on the level of access an attacker requires to carry out a side-channel attack successfully. An attacker might be working on the same system; another scenario explains the attacker's presence in the vicinity of the victim system, and the last aspect considers that the attacker is working remotely and does not require physical access to the system or its proximity. 
The last parameter for this classification is \textit{Component}, which points to the unit the attacker used. It could be a hardware unit or a part of a software unit. Mostly, side-channel attackers used hardware units in the initial days, but it requires an attacker to be physically present with the system or its vicinity. Due to this constraint, remote side-channel attacks are a new interest of attackers.

\subsection{Based on the exploited characteristic}
As already discussed, \textit{exploited characteristics} indicates that a measurable characteristic shows a unique behavior with the ongoing operation or the data involved. Further classification of side-channel attacks based on exploited characteristics derives five following types: 

\subsubsection{Timing Side-Channels}
These attacks exploit the time a device takes to perform certain operations to extract sensitive information. The computation factor and time depend on several parameters of the system. 
Timing side-channel attacks can extract information such as cryptographic keys, passwords, and other confidential data. Each process or operation has a unique timing characteristic we learned from \cite{ref_kocher_dpa}.
The success of this attack depends on the precise timing measurement that is challenging to compute, which makes these attacks hard to carry out. However, these attacks are very effective as they can extract information that is impossible or difficult to extract using other attacking methods. 
Several timing side-channel attacks came to notice on various platforms.
Examples of timing side-channel attacks are cache timing attacks, branch timing attacks, meltdown \cite{researcher16, tab13}, spectre \cite{tab12, tab17} etc.

\begin{table}[h]
    \centering
    \resizebox{\linewidth}{!}{
    \begin{tabular}{cl}
    \toprule[0.6mm]
         \textbf{Type of SCA} & \textbf{Attack} \\ \addlinespace[0.2em] \toprule 
         \multirow{6}{*}{Timing SCA} & Meltdown Attack\cite{researcher16,tab13} \\   \cmidrule(lr){2-2} 
         &Spectre Attack\cite{tab12,tab17} \\   \cmidrule(lr){2-2}
         &Rowhammer Attack\cite{rowhammer_2_rambleed,rowhammer,rowhammer1}\\   \cmidrule(lr){2-2}
         &Portsmash Attack\cite{port_contention}\\  \cmidrule(lr){2-2}
         &Cachebleed Attack\cite{cachebleed}\\   \cmidrule(lr){2-2}
         &Lucky Thirteen Attack\cite{lucky_thirteen,lucky_thirteen1}\\ \midrule
         
         \multirow{7}{*}{Power SCA} & Simple Power Analysis\cite{researcher7} \\    \cmidrule(lr){2-2}
         &Differential Power Analysis\cite{ref_kocher_dpa} \\   \cmidrule(lr){2-2}
         &High-Order DPA\cite{high_order1,high_order2,high_order}\\   \cmidrule(lr){2-2}
         &Correlation Power Analysis\cite{correlation2,correlation3,correlation1}\\   \cmidrule(lr){2-2}
         &Template Power Analysis\cite{template_power_em,template_power}\\   \cmidrule(lr){2-2}
         &Zero Power Attack\\   \cmidrule(lr){2-2}
         &Power Glitch Attack\cite{power_glitch3,power_glitch,poewr_glitch2,power_glitch1}\\ \midrule
         
         \multirow{6}{*}{EM SCA} & Simple EM Analysis\cite{simple_em_generic,simple_em,simple_em1} \\   \cmidrule(lr){2-2} 
         &Differential EM Analysis\cite{differential_em,differential_em2,differential_em1} \\   \cmidrule(lr){2-2}
         &Correlation EM Analysis\cite{correlation_em2,correlation_em,correlation_em1}\\   \cmidrule(lr){2-2}
         &High-frequency EM Analysis\\   \cmidrule(lr){2-2}
         &Zero-Power EM Attack\\   \cmidrule(lr){2-2}
         &EM Fault Injection Attack\cite{em_fault_injection,em_fault_injection1}\\ \midrule
         
         \multirow{4}{*}{Acoustic SCA} & Acoustic Cryptanalysis\cite{acoustic_cryp1,ref_sca_web3} \\    \cmidrule(lr){2-2}
         &Acoustic Attack on Printer\cite{acoustic_print2,acoustic_print1}\\    \cmidrule(lr){2-2}
         &Acoustic Attack on Hard Drive\cite{acoustic_hdd1,acoustic_hdd2,acoustic_hdd3}\\   \cmidrule(lr){2-2}
         &Acoustic Attack on Mobile\cite{acoustic_mobile2,acoustic_mobile3,acoustic_mobile4,acoustic_mobile1,acoustic_mobile5}\\ 
        \midrule
        
        \multirow{6}{*}{Optical SCA} & Optical Emanations of Display\cite{opt_em_disp,opt_em_disp1} \\   \cmidrule(lr){2-2}
         &Optical Attack on Mobiles\cite{opt_mobile_key} \\   \cmidrule(lr){2-2}
         &Optical Attack on Keyboard\cite{opt_keyboard,opt_mobile_key}\\   \cmidrule(lr){2-2}
         &Optical Attack on IoT Devices\\   \cmidrule(lr){2-2}
         &Optical Attack on Tempest\cite{opt_em_disp1}\\   \cmidrule(lr){2-2}
         &Optical Cryptanalysis\cite{opt_crypt,opt_crypt1}\\ \bottomrule[0.6mm]
    \end{tabular}}
    \caption{List of side-channel attacks based on exploited characteristic or channel}
    \label{tab:exploited_char_table}
\end{table}

\begin{figure*}[h]
    \centering
    \includegraphics[width=\textwidth]{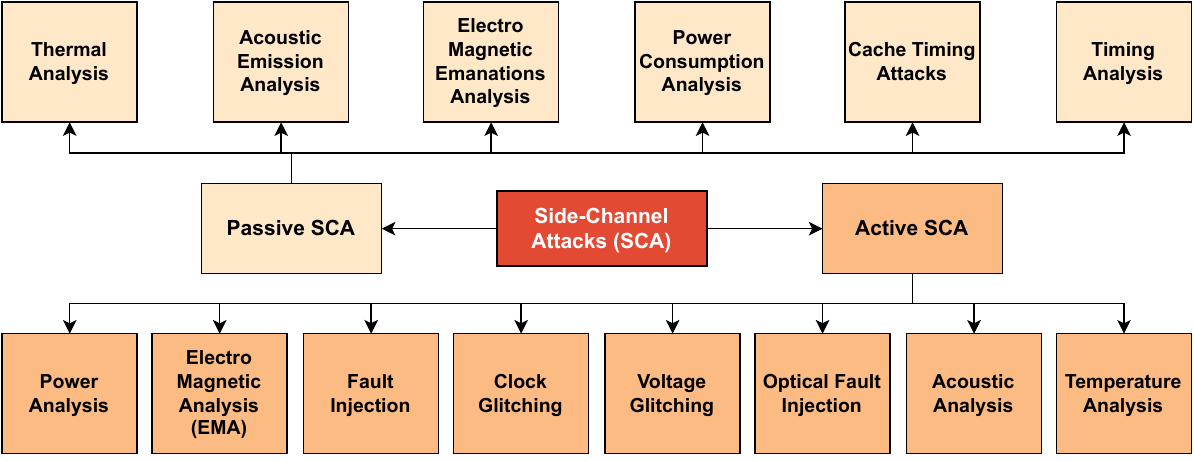}
    \caption{Active and Passive SCA Classification}
    \label{fig:active-passive}
\end{figure*}

\subsubsection{Power Side-Channels}
Information processed on a device also shows a behavioral relation with the power consumption of a computing machine. This fact leads us to power side-channel attacks that can extract sensitive information by exploiting power consumption statistics. Power side-channel attacks are not easy to deploy as they require expertise, dedicated equipment, and physical access to the system for an extended period to gain a heuristic approach. Nowadays, software-based power side-channel attacks are also in consideration. Power monitoring software access is required for attackers to work on software-based power side-channel attacks. Many attackers used Gem5 (an open-source computer system simulator) tool \cite{gem5_1, rowhammer_3} to train their model for power side-channel attacks. Similarly, Lipp et al. \cite{platypus} explored Intel's Running Average Power Limit (RAPL) library for software-based power side-channel attacks. They explained Simple Power Analysis (SPA), Differential Power Analysis (DPA), and Correlation Power Analysis (CPA) very well.  
 
\subsubsection{Electromagnetic Side-Channels}
Most computing machines emit electromagnetic radiation while performing some tasks. Researchers noticed that emitted EM radiation is not the same all the time; it differs with the operations or the data, which leads to a correlation between the emitted EMs and the operation or data involved in the process. This correlation makes the machines vulnerable to EM side-channel attacks. In section \ref{sec:side_channel_attack_history}, we discussed a document released by the NSA related to EM radiation side-channel attack in World War II. Later, in 1985, Wim van Eck \cite{ref_EMradiation} demonstrated an EM emission attack to recreate video signals from a video display unit by eavesdropping. Like power side-channel attacks, EM side-channel attacks require specialized equipment to measure the emitted radiations. Such requirements make these attack's deployment harder. EM side-channel attacks were used in hacking smart cards and ATMs \cite{smart_card2,smart_card3,smart_card1}, stealing cryptography keys \cite{em_on_cryptography1,em_on_cryptography3,em_on_cryptography2,em_on_cryptography4,em_on_cryptography5}, spying on electronic devices \cite{em_on_spying_devices1,em_on_spying_devices2,em_on_spying_devices4,em_on_spying_devices3}, etc.  
\subsubsection{Acoustic Side-Channels}
A side-channel attack that exploits the sound emissions to gain sensitive information is categorized under acoustic side-channel attack. It is noticed that computing machines or their hardware units make specific noise or sound while performing particular operations. This also requires some dedicated equipment to process the sound emission and profile it in the form of information. In 2004, Shamir et al. presented a talk (later, published an article on it in 2017 \cite{acoustic_cryp1}) titled ``Acoustic cryptanalysis: On nosy people and noisy machines'' at Eurocrypt 2004 \cite{acoustic_talk} in which they raised various aspects of acoustic side-channel attacks. Based on their conducted experiment, they suggested that in most computing devices, every operation has an acoustic signature -- a characteristic sound. The vibration of transistors causes this sound, the flow of current through wires, and other electrical activity. In 2005, researchers processed audio recordings of typed text to recover the text at UC Berkley \cite{acoustic_berkeley2005}. Primarily, acoustic side-channel attacks targeted keyboard typing \cite{acoustic_KB_1, acoustic_KB_3, acoustic_KB_2}; later, cellular devices came into existence, and then it became a common practice to use acoustic side-channel attacks \cite{acoustic_mobile2,acoustic_mobile3,acoustic_mobile4,acoustic_mobile1,acoustic_mobile5,acoustic_mobile6}. 
\subsubsection{Optical Side-Channels}
Optical side-channel attacks exploit optical emission to extract sensitive information. Kuhn \cite{optical_kuhn} analyzes the light emitted from a CRT screen to eavesdrop on the content of the display. Kuhn reconstructed readable text by deconvolving the received signal with the help of a fast photosensor. In 2015, Wang et al. \cite{optical_singlechip} discussed active and passive optical side-channel attacks on cipher chips in detail. The authors performed an optical fault injection attack on an AT89C52 singlechip in this article to gain secret information. A list of phenomenal works on optical side-channel attack is in TABLE \ref{tab:exploited_char_table}, which lists some real-life side-channel attacks classified based on exploited characteristics or channels.

\subsection{Based on the device modification}
The previous section mentioned that side-channel attacks are challenging to implement several times. The effective reasoning behind this is the measurement of correlated characteristics with the desired information, and profiling the measures requires additional hardware and time, leading us to active side-channel attacks. Active SCA induces changes in the behavior (it might be physical or logical) of the target system, while passive SCA does not require such interference. The side-channel attacks can be categorized under this classification. Fig. \ref{fig:active-passive} shows different attacks based on active and passive side-channel attacks.

\subsubsection{Active Attacks}
Active side-channel attacks intentionally induce controlled perturbations or faults into a target system to extract sensitive information. This contrasts with passive side-channel attacks, which simply observe the system's unintended side-channel emissions. It sometimes requires physical access to the victim’s system, and sometimes it can be done remotely. Active side-channel attacks combine observation and interference. By modifying inputs, clock frequencies, or environmental conditions, attackers target to strengthen and exploit side-channel leakage, such as power consumption, electromagnetic radiation, optical emissions, etc. This leakage can then be used to reveal cryptographic keys or any other confidential data.

Active side-channel attacks offer several advantages over passive attacks:
\begin{itemize}
    \item Overcome noise in the side-channel signal.
    \item Amplify weak side-channel signals.
    \item Bypass some countermeasures designed to mitigate passive side-channel attacks.
\end{itemize}

\subsubsection{Passive Attacks}
These attacks do not include manipulating, modifying, or destroying the victim’s system or execution environment. It gains unintended information by observing victim system and exploits unintended information leakage from a target system, such as variations in power consumption, electromagnetic emissions, timing patterns, etc. 

The primary purpose of these side-channel attack is to gain leaked characteristics and sensitive data, like cryptographic keys, without directly compromising the system's security mechanisms. Passive attacks are typically stealthy, as they do not actively tamper with the target, making them difficult to detect.Passive side-channel attacks can be used against various devices, from smart cards to computer systems. Countermeasures, such as noise reduction and algorithmic protections, are essential for mitigating the risks associated with passive side-channel attacks.

\begin{figure}[ht]
    \centering
    \includegraphics[width=\linewidth]{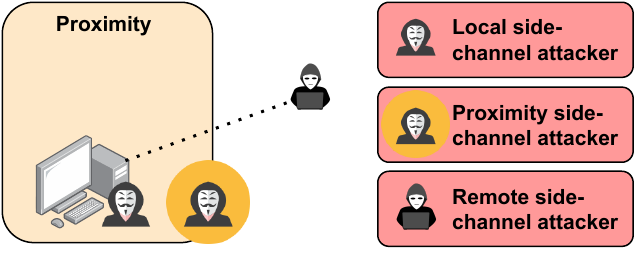}
    \caption{Access-based side-channel attackers}
    \label{fig:access-level}
\end{figure}

\subsection{Based on the access level}
This classification category depends on the attacker's physical proximity to the target. 
\subsubsection{Local attacks}
Local side-channel attacks are a type of side-channel attack performed by an attacker with physical access to the target device. This type of attack is often more challenging to defend against than remote side-channel attacks, as the attacker has more control over the environment in which the attack is performed. Local SCA include crucial characteristics measurement, which requires the victim systems access like power analysis, timing analysis, electromagnetic analysis, etc. 
\subsubsection{Proximity attacks}
Proximity side-channel attacks are performed by an attacker close to the target device but do not have physical access to it. This type of attack is often easier to defend against than local side-channel attacks, but it can still be challenging to mitigate. Proximity SCA include the measurement of characteristics that can be measured with a distance, like acoustic analysis, optical analysis, environmental tempering, etc.
\subsubsection{Remote attacks}
Remote side-channel attacks are performed by an attacker far away from the target device. This type of attack is often the most difficult to defend against, as the attacker does not have much control over the environment in which the attack is performed. Remote SCA targets the measurement of characteristic which does not require attacker and victim in the same vicinity, like cache timing analysis, fault analysis, traffic analysis, etc. 

Fig. \ref{fig:access-level} shows different access proximity parameters of local, proximity, and remote side-channel attackers. 

\begin{figure*}[b]
    \centering
    \includegraphics[width=\textwidth]{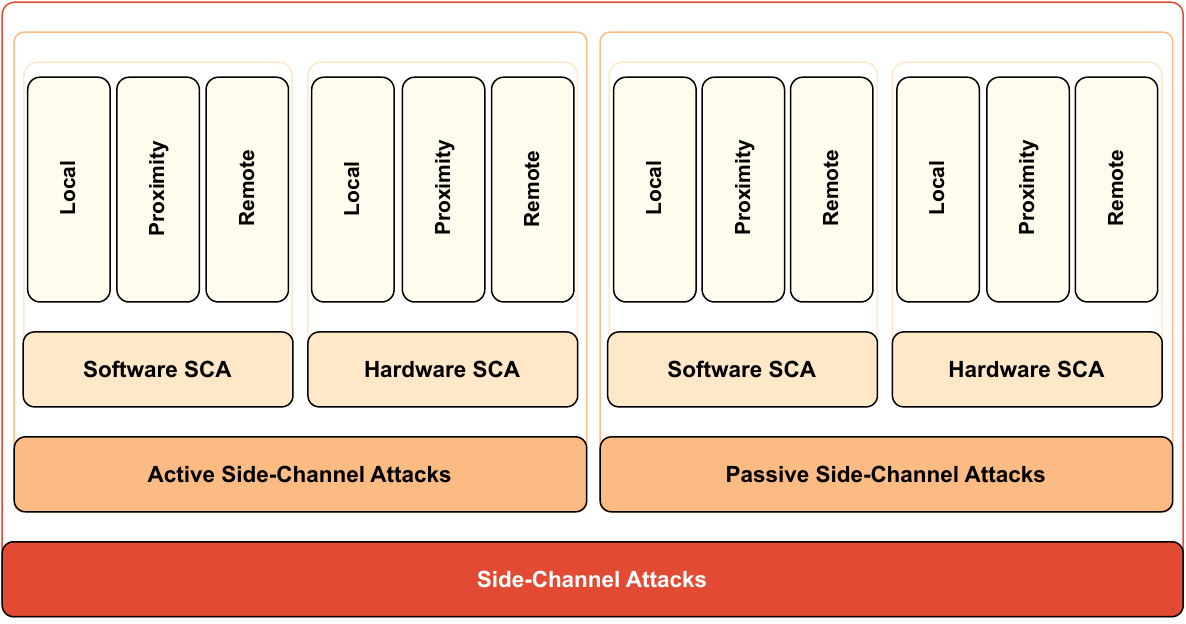}
    \caption{Overall side-channel taxonomy}
    \label{fig:overall-taxonomy}
\end{figure*}

\subsection{Based on the component }
This classification depends on the component that shows the characteristics of the correlation with sensitive information. The attack is listed under hardware and software side-channel attacks depending upon the component's type. 

\subsubsection{Hardware attacks}
Hardware side-channel attacks exploit the physical implementation of an algorithm to extract sensitive information, such as encryption keys and passwords. These attacks measure a device's power consumption, timing, or electromagnetic emissions while performing a cryptographic operation.

\subsubsection{Software attacks}
Software side-channel attacks exploit how specific algorithms are implemented in software to extract sensitive information. Software-based side-channel attacks analyze the cache usage, memory access patterns, or timing of the software while performing a task.

\subsection{Taxonomic Analysis}
We categorized side-channel attacks on various parameters into four classes. These classes are currently independent, and there is still a scope to analyze them as a taxonomical unit. It is essential to distinguish between different attack types without relying on specific data points and to explore potential connections among them. The previous section explains one fundamental division between active and passive side-channel attacks. We consider the active and passive categorization as coarse-grained and fine-grained classification for further levels as hardware and software-based variants. Additionally, the physical proximity of the attacker to the target is crucial. Local or vicinity-based attacks require close physical presence, while remote attacks can be carried out from a distance. So, these attacks are the next level of this taxonomy; after considering all these levels, an attacker can exploit a target system with a specific characteristic. This taxonomical structure is one of many possibilities; the previously mentioned four classifications can be considered at any level. However, this structure is more feasible if the entire side-channel class is considered, as this considers the attacker's capability first, followed by the target component and the proximity of the system. Fig. \ref{fig:overall-taxonomy} explains the taxonomical structure we considered. Exploited characteristic is not mentioned in Fig. \ref{fig:overall-taxonomy} because it may vary depending on the path chosen by the attacker. 

\section{Cache Side-Channel Attacks}
A cache side-channel attack exploits the cache memory to leak sensitive information in modern computer systems. The idea behind such attacks is that an attacker keeps an eye on the access pattern behavior of the shared cache and infers sensitive information about the victim’s operations. The attacker exploits the fact that various processes and threads share the cache memory and the cache’s state changes by memory access patterns. It is essential to talk about how cache side-channel attacks evolved, and their assumption are the key factors that can change the entire scenario of the cache side-channel attacks. 

\subsection{How it evolved}
Initially, this started with L1 cache side-channel attacks. At that time, the attacker had only two choices: to exploit the data cache or the instruction cache as it has low storage capacity, so there was not much possibility compared to other cache memories. The most important thing was the cache-state change event; if cache-state changes occur uniquely according to various operations, it would have been easy to find the actual procedure for the changed cache-state. L1 cache side-channel had one more issue: an attacker and victim program must reside under the same physical core to obtain that SMT was important. With time and technological evolution, more stable and vast storage has been added to the cache memory. Moving towards a higher level, exploitation of LLC becomes an extended attack vector for attackers. It opened the door to exploit the outer area of the victim program. With this, the scope of information retrieval through the side-channel attack becomes wide. Later, researchers worked on various possibilities to exploit cache side-channel attacks to an endless range approaching the Root level privileges. After such attacks, researchers drove crazy about cache side-channel attacks. Many researchers were interested in compromising KASLR (Kernel Address Space Layout Randomization) and succeeded. It also led to learning keyboard strokes through the cache and acoustic side-channels. \\
\indent Cache side-channels had the limitation and constraint of using some extra modules from the attacker's side upon arrival. With time, working on such attacks does not require any additional components. This kind of attack becomes prominent because of the constraint-free nature of the attacks. Nowadays, a primary or default configuration leads to a process to leak the entities causing the side-channel behavior; that provides ease for an attacker to attack.

\subsection{Requirements or Assumptions}
Whenever any discussion occurs regarding any attack on computer security, it requires some functional relationship between the target information and the target component. So, the cache side-channel attacks are also close to these restrictions. However, every cache side-channel has its assumptions and requirements. However, some basic assumptions are necessary to complete any cache side-channel attack. The following assumptions are a must for cache side-channel attacks: 

\subsubsection{Leakage}
There exists a correlation between the secret data being processed by the system and some observable physical or software behavior, such as cache access pattern, power consumption, timing, or electromagnetic radiation.
\subsubsection{Independence}    
The observable behavior is statistically independent of the secret data being processed, except for the correlation that leaks the information.
\subsubsection{Reproducibility}    
The observable behavior can be reproduced across multiple runs of the system under the same conditions, allowing an attacker to obtain a large number of samples for statistical analysis.
\subsubsection{Accuracy}    
The observable behavior can be measured with sufficient accuracy to distinguish between different values of the secret data.
\subsubsection{Knowledge}    
The attacker has some knowledge of the system, such as its design or software implementation, that allows them to target the side-channel vulnerabilities in an effective way.

If there is no relation between the information the attacker wants to steal and the component the attacker wants to exploit, a cache-based side-channel attack can not exist. So the first condition mentioned above is the backbone of this entire cache side-channel attack. If a change in information is not mapped to the change in cache state, it will not raise any situation where the cache's state can reveal the secret information.

Now think about the second condition; if there is a relationship between target information and target component, then it is required that an outer program must evade the area of cache in which data is stored (Mapped form, in the term of cache states).

If the second condition is approved, but the attacker program can not infer information from the changes in cache states, then everything has vanished. The third, fourth,and fifth condition come into existence here. It changes all the perceptions and allows an attacker program to sense the change in the cache state, and the attacker learns its behavior and predicts the confidential data. It is unnecessary to map perceived changes to the information physically, but it will also work if a meaningful logical mapping exists.

So these are the essential conditions required to meet the attacking scenario; if the computer system has policies that do not allow these conditions, then there is a possibility to avoid such attacks.
\section{Analysis Model}

Before proceeding to the analysis of the cache side-channel attacks, all the important factors affecting the cache side-channels are discussed in this section. So, a cache side-channel attack function can be defined as the following:

\[
 CACHE\_SCA = f(V,W,X,Y,Z)
\]
\begin{table}[H]
    \centering
    \begin{tabular}{c c l}
         V&=& $vuln\_ass$ $\in$ $ CVE$ $ or$ $ (leakage)$ \\
         W&=& $arch$ $\in$ $ MICROARCHITECTURE$ \\
         X&=& $clevel$ $\in$ $CACHE\_LEVEL$   \\
         Y&=& $method$ $\in$ $ATTACK\_METHODOLOGY$ \\
         Z&=& $scope$ $ \in$ $AFFECTED\_AREA$
    \end{tabular}
\end{table}

Here, CVE or leakage plays a vital role as it is the entry point in the victim's program or system. If there is no CVE or leakage information available to the attacker, it will not be possible to proceed further to exploit the victim's system or program for the attacker.             

\subsection{Vulnerability Assessment}
In Section I, we have discussed vulnerabilities in brief. To perform cache side-channel attack analysis, one must follow the vulnerability assessment as the attack starts with the attack vector indicated by the vulnerability. Before performing a side-channel attack, the attacker must determine which component to attack to steal confidential information. CVE (Common Vulnerability and Exposures) is a database The MITRE Corporation maintains to identify, define, and catalog publicly disclosed cybersecurity vulnerabilities. This database has a list of year-wise published vulnerabilities.
\begin{table}[h]
    \centering
    \resizebox{\linewidth}{!}{
    \begin{tabular}{cp{4.6cm}}
        \toprule[0.6mm]
        \textbf{CVE Id} & \textbf{Description} \\
        \addlinespace[0.2em] \toprule
        CVE-2022-37459 & Ampere Altra devices before 1.08g and Ampere Altra Max devices before 2.05a allow attackers to control the predictions for return addresses and potentially hijack code flow to execute arbitrary code via a side-channel attack, aka a ``Retbleed'' issue.\\ \midrule
        CVE-2022-35888 & Ampere Altra and Ampere Altra Max devices through 2022-07-15 allow attacks via Hertzbleed, a power side-channel attack that extracts secret information from the CPU by correlating the power consumption with data being processed on the system.  \\ \hline
        CVE-2022-26296 & BOOM: The Berkeley Out-of-Order RISC-V Processor commit d77c2c3 was discovered to allow unauthorized disclosure of information to an attacker with local user access via a side-channel analysis.  \\ \midrule
        CVE-2022-2612 & Side-Channel information leakage in Keyboard input in Google Chrome before 104.0.5112.79 allowed a remote attacker who had compromised the renderer process to obtain potentially sensitive information from process memory via a crafted HTML page. \\ \midrule
        CVE-2022-23304 & The implementations of EAP-pwd in hostapd before 2.10 and wpa\_supplicant before 2.10 are vulnerable to side-channel attacks as a result of cache access patterns.  \\ \bottomrule[0.6mm]
    \end{tabular}}
    \caption{Recent Side-Channel Vulnerability identified by CVE}
    \label{tab:my_table_1}
\end{table}
It has vulnerabilities for various hardware platforms, environments, and applications. For example, CVE-2022-26296 is a vulnerability found in the year 2022. The description of this vulnerability states that ``BOOM: The Berkeley Out-of-Order RISC-V Processor commit d77c2c3 was discovered to allow unauthorized disclosure of information to an attacker with local user access via a side-channel analysis.'' Which directly states the definition of a new side-channel attack that can be done on the RISC-V Processors. It could lead to sensitive data leakage through side-channel attack/analysis. A list of recent side-channel vulnerabilities is given in the TABLE \ref{tab:my_table_1}.

\subsection{Micro-architecture}
Intel, AMD, and ARM are the three major players in the semiconductor industry, each with its strengths and weaknesses. Most desktops and laptops use Intel; that's why Intel is the market leader in x86 processors \cite{micro_market}. AMD is Intel's main competitor in the x86 market and has gained market share in recent years. ARM is the market leader in mobile processors and is also increasingly being used in other devices, such as servers and IoT devices. Intel, AMD, and ARM use different microarchitectures for their processors. Intel uses a complex instruction set computer (CISC) architecture, while AMD and ARM use reduced instruction set computer (RISC) architectures. CISC architectures are more powerful and versatile but also more complex and less power-efficient. RISC architectures are simpler and more power-efficient but also less powerful and versatile. AMD processors use a non-inclusive LLC, which means that the LLC is not always synchronized with the L1 and L2 caches. This makes them more resistant to Flush + Reload attacks than Intel processors, which use an inclusive LLC\cite{tab30}.

\subsection{Cache Level}
In section III, three types of cache memory were explained for different purposes L1, L2, and L3. L1 and L2 are separate for each core, and L3 is a shared cache memory between all cores. 

\begin{figure}[h]
    \centering
    \includegraphics[width=\linewidth]{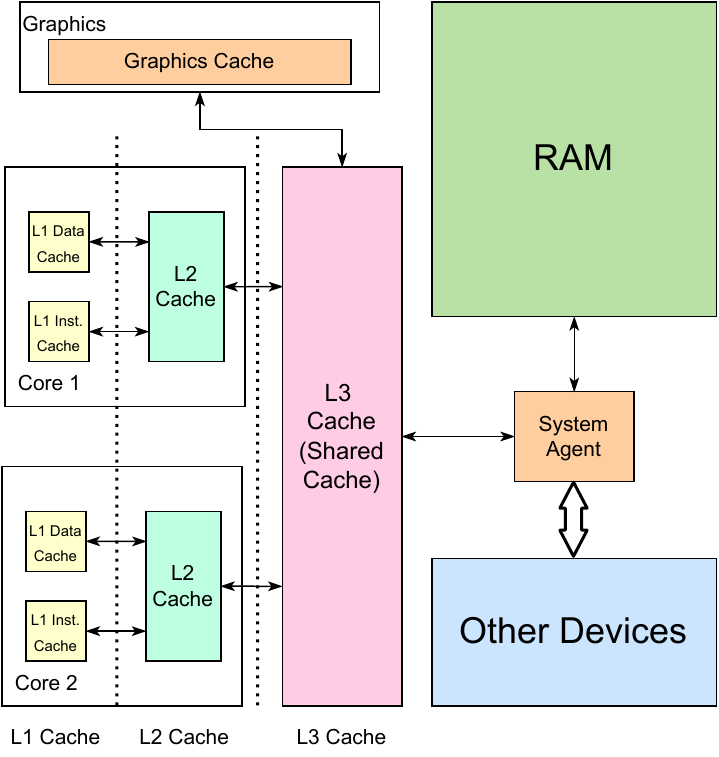}
    \caption{General System Architecture}
    \label{fig:fig_cache_architecture}
\end{figure}
In Fig. \ref{fig:fig_cache_architecture}, the dataflow and workflow of various cache memories have been illustrated in detail. A long time ago, only L3 appeared in the architecture when only single-core systems existed, but when companies introduced multi-core systems, L1 and L2 were also entertained. Cache memories are high-speed compared to RAM or any secondary storage device. L1 is further divided into two segments, L1-I and L1-D. L1-I deals with the information about the operation that the CPU has to perform. L1-D holds the data on which the operation is to be performed. L1-I, L1-D, and L2 are private caches, as each core has its own set of L1-I, L1-D, and L2 caches. L1-I and L1-D are immediate caches to the processor or CPU, so whenever any instruction executed by the processor will affect L1-I and L1-D immediately. If L1-I and L1-D have a significant impact and that impact is measurable, then stealing information will be easy through cache side-channel attacks.
Researchers\cite{tab2,tab5,tab14,tab23} used L1-D for side-channel attacks and some researchers\cite{tab21,tab6} used L1-I for side-channel attacks.
As L2 has a higher storage capacity than L1 and is shared between multiple threads/processes working on the same core, it behaves like a data carrier between L1 and L3 most of the time, so L2’s exploitation is not as promising as L1 and L3. There are fewer attacks on L2 than on L1 and L3. L3 has high storage capacity and is shared between cores; it has more scope to compromise.
Most of the researchers\cite{tab8,tab20,tab30,tab4,tab31,tab33,tab34,tab9,tab35,tab36,tab37,tab22,tab38,tab39,tab16,tab12,tab13,tab17,tab40,tab41} used L3(Last Level Cache) to carry out side-channel attack and achieved vast information.

\begin{table*}[ht]
    \centering
    \resizebox{\linewidth}{!}{
    \begin{tabular}{llcccccccc}
    \toprule[0.6mm]
        \textbf{Author(s)} & \textbf{Year} & \textbf{Intel} & \textbf{AMD} & \textbf{ARM} & \textbf{Cross\textsuperscript{$\ast$}} & \textbf{CC\textsuperscript{$\dagger$}} & \textbf{Methodology} & \textbf{Range} & \textbf{Target} \\ \addlinespace[0.2em]
\toprule
        Osvik et al. \cite{tab5}  &  2006  &   \checkmark  &  $\times$  &  $\times$  &  $\times$  &  L1-D  &  Prime + Probe  &  Core Level  &  OpenSSL (0.9.8) AES \\ \midrule
        Onur Acii\c{c}mez \cite{tab21}  &  2007  &   \checkmark  &   \checkmark  &   \checkmark  &  $\times$  &  L1-I  &  Prime + Probe  &  Core Level  &  OpenSSL (0.9.8d) RSA \\ \midrule
        Neve et al. \cite{tab2}  &  2007  &   \checkmark  &  $\times$  &  $\times$  &  $\times$  &  L1-D  &  Prime + Probe  &  Core Level  &  AES \\ \midrule
        Onur Acii\c{c}mez et al. \cite{tab6}  &  2008  &   \checkmark  &   \checkmark  &   \checkmark  &  $\times$  &  L1-I  &  Prime + Probe  &  Core Level  &  OpenSSL (0.9.8e) RSA \\ \midrule
        Colin Percival \cite{tab14}  &  2009  &   \checkmark  &   \checkmark  &  $\times$  &  $\times$  &  L1-D  &  Prime + Probe  &  Core Level  &  OpenSSL (0.9.8c) RSA \\ \midrule
        Tromer et al. \cite{tab23}  &  2010  &   \checkmark  &  $\times$  &  $\times$  &  $\times$  &  L1-D  &  Prime + Probe  &  Core Level  &  OpenSSL (0.9.8) AES \\ \midrule
        Gullasch et al. \cite{tab20}  &  2011  &   \checkmark  &  $\times$  &  $\times$  &   \checkmark  &  LLC  &  Flush  + Reload  &  Application Level  &  OpenSSL (0.9.8n) AES \\ \midrule
        Hund et al. \cite{tab8}  &  2013  &   \checkmark  &   \checkmark  &  $\times$  &   \checkmark  &  LLC  &  Evict + Time  &  Kernel Level  &  Kernel Memory \\ \midrule
        Irazoqui et al.  \cite{tab4}  &  2014  &   \checkmark  &  $\times$  &  $\times$  &   \checkmark  &  LLC  &  Flush + Reload  &  Kernel Level  &  OpenSSL (1.0.1f) AES \\ \midrule
        Yarom et al. \cite{tab34}  &  2014  &   \checkmark  &  $\times$  &  $\times$  &   \checkmark  &  LLC  &  Flush + Reload  &  Application Level  &  OpenSSL (1.0.1e) ECDSA \\ \midrule
        Yarom et al. \cite{tab30}  &  2014  &   \checkmark  &  $\times$  &  $\times$  &   \checkmark  &  LLC  &  Flush + Reload  &  Kernel Level  &  GnyPG (1.4.13) RSA \\ \midrule
        Zhang et al. \cite{tab9}  &  2014  &   \checkmark  &  $\times$  &  $\times$  &   \checkmark  &  LLC  &  Flush + Reload  &  Kernel Level VM  &  User Activities \\ \midrule
        Gruss et al. \cite{tab36}  &  2015  &   \checkmark  &  $\times$  &   \checkmark  &   \checkmark  &  LLC  &  Flush  + Reload  &  Application Level  &  Keyboard Input \\ \midrule
        Irazoqui et al.  \cite{tab35}  &  2015  &   \checkmark  &  $\times$  &  $\times$  &   \checkmark  &  LLC  &  Flush + Reload  &  Kernel Level  &  TLS and DTLS \\ \midrule
        Liu et al. \cite{tab37}  &  2015  &   \checkmark  &  $\times$  &  $\times$  &   \checkmark  &  LLC  &  Prime + Probe  &  Kernel Level  &  GnyPG (1.4.131.4.18) ElGamal \\\midrule
        Pol et al. \cite{tab33}  &  2015  &   \checkmark  &  $\times$  &  $\times$  &   \checkmark  &  LLC  &  Flush + Reload  &  Application Level  &  OpenSSL (1.0.1e) ECDSA \\ \midrule
        Allan et al. \cite{tab31}  &  2016  &   \checkmark  &  $\times$  &  $\times$  &   \checkmark  &  LLC  &  Flush + Reload  &  Application Level  &  OpenSSL (1.0.1e) ECDSA \\ \midrule
        Gruss et al. \cite{tab39}  &  2016  &   \checkmark  &  $\times$  &   \checkmark  &  $\times$  &  LLC  &  Flush + Flush  &  Core Level  &  AES T-table \\ \midrule
        Kayaalp et al.  \cite{tab38}  &  2016  &   \checkmark  &  $\times$  &  $\times$  &   \checkmark  &  LLC  &  Prime + Probe  &  Application Level  &  AES \\ \midrule
        Garc\'{\i}a et al. \cite{tab22}  &  2016  &   \checkmark  &  $\times$  &  $\times$  &   \checkmark  &  LLC  &  Flush + Reload  &  Kernel Level  &  OpenSSL DSA \\ \midrule
        Brasser et al. \cite{tab40}  &  2017  &   \checkmark  &  $\times$  &  $\times$  &   \checkmark  &  LLC  &  Prime + Probe  &  Application Level  &  RSA \\ \midrule
        Schwarz et al. \cite{tab41}  &  2017  &   \checkmark  &  $\times$  &  $\times$  &   \checkmark  &  LLC  &  Prime + Probe  &  Application Level  &  RSA \\ \midrule
        G\"{o}tzfried et al. \cite{list_1_intel_sgx}  &  2017  &   \checkmark  &  $\times$  &  $\times$  &   $\times$  &  L1  &  Prime + Probe  &  Core Level  &  OpenSSL (0.9.7a) AES \\ \midrule
        Atici et al. \cite{list_2_profiling}  &  2018  &   \checkmark  &  $\times$  &  $\times$  &   \checkmark  &  Smart Cache  &  Timing Attack  &  Application Level  &  OpenSSL (0.9.7a) AES \\ \midrule
        Briongos et al. \cite{list_3_cache_miss}  &  2019  &   \checkmark  &  $\times$  &  $\times$  &   \checkmark  &  LLC  &  FR, FF, PP  &  Application Level  &  OpenSSL (1.0.1f) AES \\ \midrule
        Moghimi et al. \cite{list_4_memjam}  &  2019  &   \checkmark  &  $\times$  &  $\times$  &   $\times$  &  L1  &  Evict + Time  &  Core Level  & DES, AES and SM4  \\ \midrule
        Wang et al. \cite{list_5_papp}  &  2019  &   \checkmark  &  $\times$  &  $\times$  &   $\times$  &  L2  &  Prime + Probe  &  Core Level  &  AES \\ 
        \bottomrule[0.6mm]
        \end{tabular}}
            \begin{tablenotes}
            \item \textsuperscript{$\ast$}Cross-Core
            \item \textsuperscript{$\dagger$}Cache-Component
    \end{tablenotes}
    \caption{Research data on cache side-channel attacks}
    \label{tab:my_label}
\end{table*}
\subsection{Attack Methodology}
\subsubsection{Prime + Probe}
The simplest attack that an attacker can do is Prime + Probe. This attack includes three stages. First, the attacker primes the cache memory, which fills the entire cache memory with its content. So now, as the attacker has filled the entire cache memory so, whenever the attacker accesses their data will take a particular time (limited period). When the victim process starts processing, some of the data is filled in the random cache lines of the cache memory. Due to this conflict, the attacker’s data from the particular cache lines will be evicted and filled with the data by the victim. \\
Now in the probing stage, when the attacker tries to access its data from different cache lines, some will take longer than the threshold time( Cache access time + Main Memory Access time). In this scenario, the attacker will check the time, know which cache lines have victim data, and probe the victim data from the cache memory. 
\begin{figure}
\centering
\begin{minipage}{.24\textwidth}
  \centering
  \includegraphics[scale=0.55]{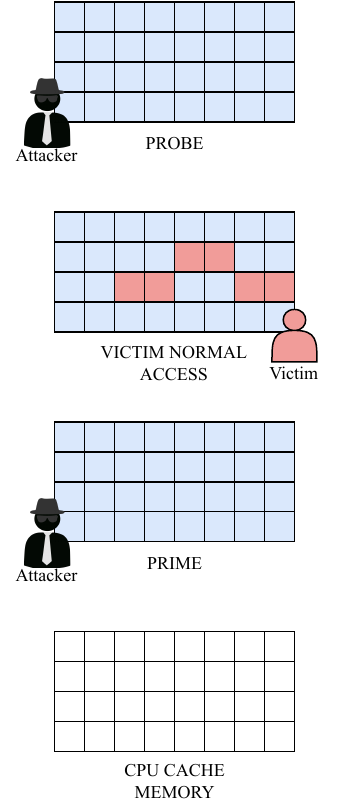}
  \caption{Prime and Probe}
  \label{fig:test1}
\end{minipage}%
\begin{minipage}{.24\textwidth}
  \centering
  \includegraphics[scale=0.55]{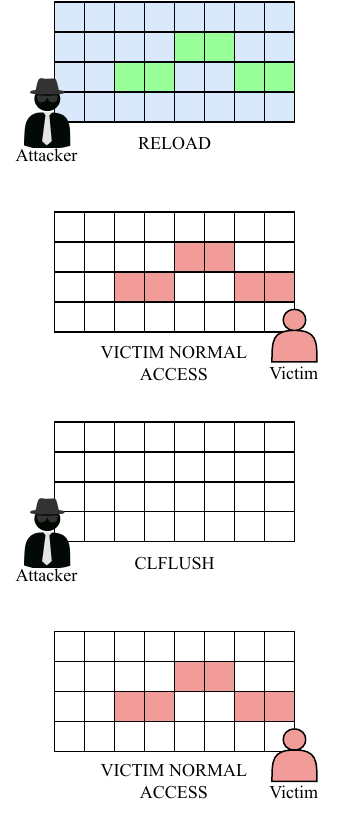}
  \caption{Flush and Reload}
  \label{fig:test2}
\end{minipage}
\end{figure}
\subsubsection{Flush + Reload}
This attack is a variant of the previous attack. The content is available in the L1 (High-Level Cache) cache and will always be available in L3(LLC: Last Level Cache). Flush + Reload exploits this multilayer inheritance structure of the Cache Memory. When any content is evicted from the high-level cache, it will also be evicted from the low-level cache. To flush cache content, the attacker uses the CLFLUSH instruction. When the victim processes the usage cache and fills cache lines with its content, the attacker then uses the CLFLUSH instruction to empty the cache memory, wipes all the content available on the cache memory, and then sits idle. When the victim processes usages of the cache memory again and the attacker, after a specified period, checks whether victim content is available on the cache lines. If it is available, it can be concluded that the victim accessed particular cache lines, and its data is available there. The attacker then reloads the entire cache memory content, and now the attacker knows the content used by the victim process. This attack will work because the attacker and victim process will have a shared memory. If this is not true, it will not work further.
\subsubsection{Evict + Reload}
This attack is similar to the previous Flush + Reload attack. The eviction method plays a critical role here in separating both attack modules. As discussed in Flush + Reload, it will only work if shared memory is available; otherwise, it will not work. If the shared memory is available there, it can evict the corresponding cache lines directly by the CLFLUSH instruction; otherwise, it will not work like this. So in Evict + Reload, no shared memory is available, so we can not use the CLFLUSH instruction. It exploits the eviction of cache lines by filling its data. So the victim process’s data will be evicted from the cache lines. The attacker then waits for a specified period, like FLUSH + RELOAD.\\
Meanwhile, the victim process will reaccess its data and sit on some cache lines in this specified time. Now, the attacker will check if some different data is available there, and this difference can conclude that this new data belongs to the victim process. The attacker will reload the entire data from the cache memory and pick the specific changes in the cache line data. This is how it establishes a connection between the attacker and victim process.
\subsubsection{Evict + Time}
This attack does not involve any extra methods to exploit the cache memory; the attacker uses the cache memory usually, but it measures the time to establish a baseline connection. The attacker lets the victim process use cache memory first; after some time, it evicts some particular cache lines and measures the time again. This attack involves multiple execution cycles. Later by comparing the recorded time and the baseline of the victim, it tries to exploit the cache memory. However, this attack targets particular cache lines only. It can not provide efficient results just after the victim program executes once. 
\subsubsection{Flush + Flush}
In this attack, CLFLUSH instruction’s execution plays a vital role as it will show different execution times based on the content available in the cache memory. Various cache states will take different execution times. Here CLFLUSH will target some cache line data, and if it is available there, then multilevel cache eviction during the execution will be required to make execution time longer; otherwise, it will get a short execution time if it does not evict the multilevel cache. The attacker’s program will determine the state of cache lines and whether it was used by the victim program based on the recorded time difference.\\
Once the attacker does the flush using CLFLUSH, it will allow the victim to load the content normally. Once done (by the time measurement), the attacker will again empty the cache lines using the CLFLUSH instruction and record the time. Unlike the other attacking modes, this works entirely on the execution time of various instructions rather than accessing the memory directly; it creates very high chances of false positive and negative results.
\subsection{Scope}
Scope defines the level of exploitation in cache side-channel attacks. Depending on the architecture, L1, L2, and L3 have different execution scopes. As in Fig. \ref{fig:fig_cache_architecture}, it is clear that the L1 cache is pure physical core's memory, so whatever cache side-channel attack on the L1 memory will lead to the exploitation of that core only. So it is necessary to exploit the L1 cache that attacker and victim programs must reside on the same physical core; otherwise, it will not be possible. 
All the available cores share the LLC in a computer system, and attacking LLC can lead to multiple levels of exploitation.\\
Overall these scopes can be categorized into three levels:
\begin{itemize}
    \item Core Level 
    \item Application Level
    \item Kernel Level
\end{itemize}
\subsubsection{Core Level} 
In most cases of L1-I and L1-D, it occurs where an attacker can reach the corresponding cache memory only. Core-level exploitation is helpful in cryptanalysis through side-channel attacks. Cryptographic algorithms use key(s) to perform encryption or decryption. Depending on the algorithm, some calculations are performed by the computer repetitively; in that case, cache memory serves several hits/misses. Measuring these hits/misses and access time may leak the cryptographic key, and several instances are available. 
\subsubsection{Application Level}
This level of exploitation generally happens between an application's shared objects. An application might have two views; the victim uses one, and the attacker uses another. Though both views have some shared objects, the attacker can intrude in the victim's area using the attack methodologies mentioned above since application confidentiality may be compromised; we are naming it as application level. The attacker can not access data outside the targeted application. 
\subsubsection{Kernel Levels}
This is the highest level of exploitation an attacker can achieve. In this scope, the attacker reaches kernel-level data and attains permission to work on the system's filesystem. When an attacker exploits cache for this level, it is likely the Last Level Cache(LLC), as the rest are core-private cache memories. If an attacker reaches this level of exploitation, he/she might be able to run scripts or commands affecting the entire system. In other words, it is considered that the attacker has root control over the system's terminal, but this usually comes with conditions; the attacker might require a robust data access policy through the side-channel attack or some other instances. Side-Channel comes with preassumed conditions; in this case, the victim might need the root privileges.

\section{Mitigation Techniques}
As we have seen, the cache side-channel attacks grasp the execution pattern of the cache memory hierarchy to extract sensitive information by exploiting variations in cache access times. That information might be cryptographic keys or private data. To stop or defend against these kinds of attacks, we require some methodologies against cache-based side-channel attacks. 
Though this is enough argument in favor of mitigation strategies, we can list the needful key points in 
favor of the mitigation strategies as follow.
\begin{itemize}
    \item \textit{Information Leakage:} Attackers can access unauthorized information using cache side-channel attacks.
    \item \textit{Covert Communication Channels:} An attacker can exploit these attacks as covert communication channels to transmit data covertly between two or more interdependent malicious processes.
    \item \textit{Threat to Multi-Tenant Environment:} In a virtualized environment (especially in cloud computing), cache side-channel attack poses a substantial threat. The attacker (s) can gain access or information from a VM to another co-located VM by exploiting it.
    \item \textit{Exploitation of Microarchitectural Features:} Cache side-channel attacks exploit the microarchitecture features and shared resources of modern processors. As processors grow and become more complex, new vulnerabilities and attack techniques may be discovered, highlighting the need for effective mitigation.
\end{itemize}
Cache side-channel attacks are not only because of software implementation; microarchitectural hardware designs are also responsible. So  it is essential to explore hardware and software-based mitigation techniques to address the threat posed by cache side-channel attacks. Hardware-based mitigations focus on modifying the microarchitecture or cache design to minimize the leakage of information. Techniques such as cache partitioning, where different security levels are assigned to specific cache portions, help isolate sensitive data from potential leakage channels. Additionally, hardware-assisted techniques like cache obfuscation and dynamic cache resizing introduce randomness or dynamically adjust cache sizes to impede side-channel attacks. On the other hand, software-based mitigations involve modifying software algorithms and implementations to reduce the leakage of sensitive data through cache side channels. Countermeasures like algorithmic improvements, data encryption, and data-dependent memory access patterns can significantly decrease the effectiveness of cache side-channel attacks. Considering and evaluating these hardware and software-based mitigation strategies can enhance the security of computer systems against cache side-channel attacks, thereby ensuring the confidentiality and integrity of sensitive information in the face of evolving cyber threats.
\begin{figure}
    \centering
    \includegraphics[width=\linewidth]{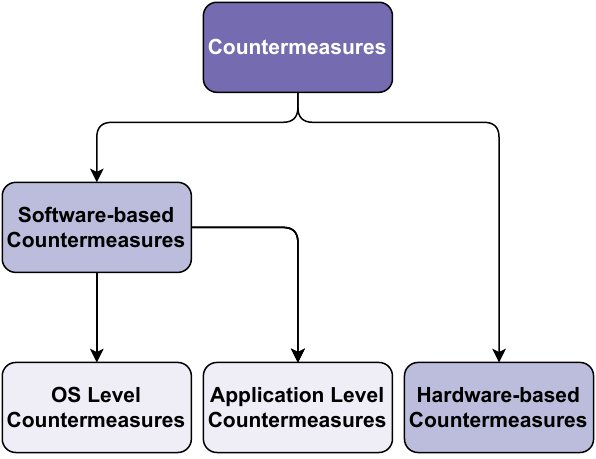}
    \caption{Hierarchy of countermeasures}
    \label{fig:hira}
\end{figure}

\subsection{Software-based Solutions}
Software-based defenses are essential in preventing cache side-channel attacks. It works on attacks that use software attack vectors. Programming techniques and constant-time algorithms are one efficient defense. Constant-time programming shields an algorithm from timing or cache-based side channels by guaranteeing that an algorithm's execution time and memory accesses are independent of secret information. Programmers can do this by carefully examining the code's control flow and data dependencies and removing any conditional branching or memory accesses that might reveal information due to cache behavior. Using cache obliviousness is another software-based defense mechanism. With this method, algorithms can be created that do not rely on any particular cache size, cache line size, or cache associativity. 
Cache obliviousness reduces the possibility of cache-based side-channel attacks by guaranteeing the algorithm's performance is constant across various cache configurations. Software may also make use of cache flushing or cache access randomization methods. After processing sensitive data, cache lines must be manually cleared or invalidated to prevent information from leaking during subsequent access. Cache access randomization, on the other hand, adds unusual noise to cache access patterns, making it more challenging for attackers to extract useful information. To construct a layered defense against cache-based side-channel attacks, software-based countermeasures can be added to current security features like encryption and access control systems. Overall, software-based countermeasures offer crucial methods to improve computer system security and shield sensitive data from side-channel attacks that rely on caches.
Further, these countermeasures can be divided into various categories as follow.
\subsubsection{Application Level Countermeasures}
Application-level countermeasures of cache side-channel attacks protect software from cache side-channel attacks without needing hardware or system-level changes. 
    \begin{itemize}
        \item Execution time unification
        \item Execution time randomization
        \item Software vulnerability detection
    \end{itemize}
\subsubsection{System Level Countermeasures} 
    \begin{itemize}
        \item Process execution partitioning
        \item Process scheduling
        \item Measurement randomization
        \item Attack detection
    \end{itemize}

\subsection{Hardware-based Solutions}
In order to guard against cache-based side-channel attacks and to provide an extra layer of security to computer systems, hardware-based countermeasures are also necessary. Implementing cache access isolation methods is a crucial countermeasure. These techniques isolate several processes or security domains to prevent unauthorized access to cache memory. Cache access isolation reduces the risk of side-channel attacks by rigorously regulating cache access privileges and ensuring that only authorized entities can access sensitive information. Cache partitioning techniques are another hardware-based defense strategy. This involves setting up multiple partitions within the cache and allocating particular partitions to various processes or security domains. Cache partitioning reduces the likelihood of information leaking through cache-based side channels by separating the cache among various entities. Cache replacement rules that reduce side-channel vulnerabilities can also be used as hardware-based countermeasures.These rules make it more difficult for attackers to infer useful data from cache behavior using cache replacement algorithms that prioritize data eviction based on variables unrelated to sensitive information, such as access time or recurrence. Furthermore, dynamic cache resizing algorithms may be incorporated as part of hardware-based countermeasures. These mechanisms alter the cache size according to the security demands of the running code, giving vital or sensitive data additional cache space while minimizing the cache footprint of potentially vulnerable processes. To build a strong defense against cache-based side-channel assaults, these hardware-based remedies can be used in concert with software-based strategies. 
\begin{itemize}
    \item Hardware resource partitioning
	\item Resource usage randomization
	\item Hardware vulnerability identification
\end{itemize}

\section{Conclusion}
This comprehensive survey paper examined the landscape of cache side-channel attacks, providing an in-depth understanding of their fundamental principles, exploitation techniques, and mitigation strategies. Our analysis highlighted the significant security risks these attacks pose and emphasized the importance of robust defenses to protect sensitive information in modern computer systems.
Cache side-channel attacks exploit the observable behavior of shared cache resources, allowing attackers to infer privileged information by leveraging timing discrepancies and access patterns. The emergence of shared cache architectures in processors, aimed at enhancing performance, inadvertently introduced a covert channel that adversaries can exploit. Consequently, understanding the mechanisms behind these attacks is crucial for developing effective countermeasures.
We explored various cache side-channel attack scenarios, including Prime+Probe, Flush+Reload, and Evict+Time, providing insights into their working principles and real-world implications. These attack scenarios demonstrated the potential risks associated with cache-side channels and highlighted the need for proactive defense mechanisms.
To mitigate cache side-channel attacks, we discussed a range of hardware-based and software-based countermeasures. Hardware-based solutions, such as cache partitioning and cache obfuscation, aim to isolate cache resources and introduce confusion to frustrate attackers. Software-based approaches, including compiler optimizations, code randomization, and cryptographic techniques, focus on minimizing information leakage and improving system resilience.
However, the implementation of countermeasures is challenging. Countermeasures can introduce performance overhead, increase energy consumption, and complicate software development processes. Achieving a balance between security and system efficiency is crucial to ensure the practicality and sustainability of mitigation strategies.
Furthermore, we considered the evolving landscape of cache side-channel attacks and identified emerging attack vectors and potential future research directions. By staying abreast of the latest trends and vulnerabilities, researchers and practitioners can proactively develop more robust defense mechanisms and anticipate the evolving nature of these attacks.
In conclusion, cache side-channel attacks pose a significant threat to the security of modern computer systems. Through this comprehensive survey, we have provided a holistic understanding of these attacks, covering their fundamental principles, exploitation techniques, and mitigation strategies. By shedding light on the potential risks and discussing effective countermeasures, we aim to empower researchers and practitioners to develop secure and resilient systems that can withstand cache side-channel attacks without compromising performance or usability.
As cache side-channel attacks continue to evolve, ongoing research and collaboration are crucial to stay ahead of attackers. By combining our collective knowledge and expertise, we can strive towards a future where cache side-channel attacks are effectively mitigated, and the confidentiality and integrity of sensitive information are safeguarded in modern computer systems.

\printbibliography
\end{document}